\documentclass[universe,review,accept,pdftex,moreauthors]{Definitions/mdpi}
\usepackage{amsmath}
\usepackage{amssymb}
\usepackage{graphicx}
\usepackage[utf8]{inputenc}
\usepackage{adjustbox}
\usepackage{natbib}

\usepackage[utf8]{inputenc}

\firstpage{1}
\pubvolume{11}
\issuenum{7}
\articlenumber{229}
\pubyear{2025}
\copyrightyear{2025}
\externaleditor{Giuseppe Verde} 
\datereceived{5 May 2025}
\daterevised{9 June 2025} 
\dateaccepted{7 July 2025}
\datepublished{11 July 2025}
\hreflink{https://doi.org/10.3390/\\universe11070229}

\def\hl{}
\def\highlighting{}

\Title{Recent Advances in Understanding $R$-Process Nucleosynthesis in Metal-Poor Stars and Stellar Systems}

\TitleCitation{Recent Advances in Understanding $R$-Process Nucleosynthesis in Metal-Poor Stars and Stellar Systems}

\Author{\hl{Avrajit Bandyopadhyay} 
$^{1,}$*$^{,\dagger}$\href{https://orcid.org/0000-0002-8304-5444}{\orcidicon} and Timothy C\hl{.} 
Beers $^{2,\dagger}$\href{https://orcid.org/0000-0003-4573-6233}{\orcidicon}}

\AuthorNames{Avrajit Bandyopadhyay and Timothy C Beers}

\isAPAStyle{%
\AuthorCitation{Bandyopadhyay, A., \& Beers, Timothy C.}
}{%
\isChicagoStyle{%
\AuthorCitation{Bandyopadhyay, Avrajit, and Beers, Timothy C.}
}{
\AuthorCitation{Bandyopadhyay, A.; Beers, T.C.}
}
}

\address{%
$^{1}$ \quad Department of Astronomy, \hl{University of Florida, Gainesville, FL 32611, USA} 
\\
$^{2}$ \quad Department of Physics and Astronomy, \hl{University of Notre Dame, Notre Dame, IN 46556, USA}; tbeers@nd.edu}

\corres{Correspondence: avrajit.ban@gmail.com or \hl{abandyopadhyay@ufl.edu} 
}

\firstnote{These authors contributed equally to this work.}

\abstract{The rapid neutron-capture process (\highlighting{$r$}
-process) is responsible for the creation of roughly half of the elements heavier than iron, including precious metals like silver, gold, and platinum, as well as radioactive elements such as thorium and uranium. Despite its importance, the nature of the astrophysical sites where the $r$-process occurs, and the detailed mechanisms of its formation, remain elusive. The key to resolving these mysteries lies in the study of chemical signatures preserved in ancient, metal-poor stars. These stars, which formed in the early Universe, retain the chemical fingerprints of early nucleosynthetic events and offer a unique opportunity to trace the origins of $r$-process elements in the early Galaxy. In this review, we explore the state-of-the-art understanding of $r$-process nucleosynthesis, focusing on the sites, progenitors, and formation mechanisms. We discuss the role of potential astrophysical sites such as neutron star mergers, core-collapse supernovae, magneto-rotational supernovae, and collapsars, that can play a key role in producing the heavy elements. We also highlight the importance of studying these signatures through high-resolution spectroscopic surveys, stellar archaeology, and multi-messenger astronomy. Recent advancements, such as the gravitational wave event GW170817 and detection of the $r$-process in the ejecta of its associated kilonovae, have established neutron star mergers as one of the confirmed sites. However, questions remain regarding whether they are the only sites that could have contributed in early epochs or if additional sources are needed to explain the signatures of $r$-process found in the oldest stars. Additionally, there are strong indications pointing towards additional sources of $r$-process-rich nuclei in the context of Galactic evolutionary timescales. These are several of the outstanding questions that led to the formation of collaborative efforts such as the $R$-Process Alliance, which aims to consolidate observational data, modeling techniques, and theoretical frameworks to derive better constraints on deciphering the astrophysical sites and timescales of $r$-process enrichment in the Galaxy. This review summarizes what has been learned so far, the challenges that remain, and the exciting prospects for future discoveries. The increasing synergy between observational facilities, computational models, and large-scale surveys is poised to transform our understanding of $r$-process nucleosynthesis in the coming years.}


\begin{document}


\section{Introduction}

Ancient, metal-poor stars serve as ``cosmic fossils'', as~they are formed primarily within the first few billion years following the Big Bang. They are low mass, long-lived, and~retain detailed records of their past host environments, allowing us to reconstruct the chemical evolution of the Galaxy~\citep{beers85,beers92,mcwilliam1995,beers2005,bromm2009,frebelandnorris}. {\hl{These} 
metal-poor stars (stars with lower content of metals compared to the Sun; in astronomy all elements heavier than H and He are referred to as ``metals'') provide crucial insight into the nucleosynthetic processes and star-formation histories of the first generations of stars in the early Milky Way, as~they retain the chemical signatures from their progenitors.} With high-resolution spectroscopic studies of these metal-poor stars, it is possible to throw light on the possible enrichment events in the early Galaxy and its chemical~evolution.

The elemental signatures imprinted in these stars encode information about their progenitors and past nucleosynthetic events, including possible contributions from neutron star\hl{--}neutron star and neutron star\hl{--}
black hole mergers, core-collapse supernovae, and~other enrichment sources. {Over the past decade, the~advent of large-scale observational surveys (e.g., SDSS, LAMOST, GALAH, and DESI), along with advances in theoretical models and experimental data, have ushered in a renaissance in the field of Galactic Archaeology, providing unprecedented insight into the
chemical--evolutionary history of the Milky Way and its satellite galaxies~\citep{Cowan2021,kobayashi2020,Schatz2022}.} This progress has established a crucial link between nuclear astrophysics, stellar evolution, and~galaxy formation, enabling us to place stringent constraints on the physical processes driving chemical enrichment in the cosmos. In~particular, the~most metal-poor stars in the Milky Way halo\hl{---}remnants of the early Universe\hl{---}offer a unique opportunity to probe the nucleosynthetic pathways that governed the formation of the heaviest elements~\citep{christlieb2003,frebelandnorris}. These stars, largely chemically undisturbed over some \mbox{12--13 Gyr}, act as pristine records of the earliest enrichment events, preserving the chemical fingerprints of individual nucleosynthetic processes that are otherwise difficult to disentangle from studies of younger or more metal-rich stars~\citep{Starkenburg2017,bandyopadhyay}.

By studying the elemental compositions of these ancient stars, we endeavor to isolate the distinct contributions from the proposed astrophysical enrichment events. Unlike laboratory nuclear physics experiments, which cannot replicate the extreme environments required for the production of the heaviest elements, these stars serve as natural testbeds for understanding $r$-process nucleosynthesis and the chemical evolution of the Universe. Their elemental-abundance patterns provide crucial insights into how heavy elements shaped the formation and evolution of stars and galaxies and~link the present-day Milky Way to its complex accretion~history.

\begin{figure}[H]

\includegraphics[scale=0.3]{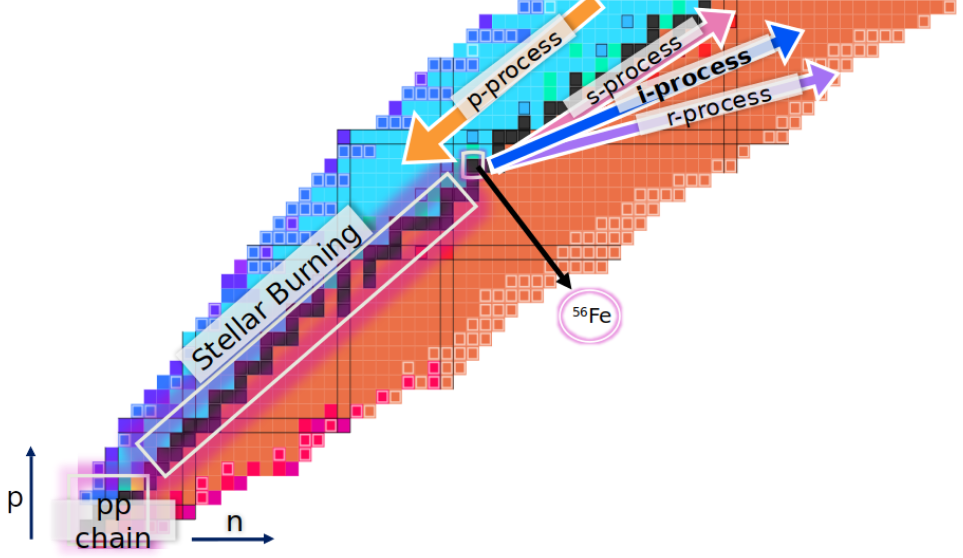}
\caption{A portion of the chart of nuclides, illustrating the different nucleosynthetic pathways in stars. Key nuclear burning processes such as the p-p chain or proton-capture reactions, and~advanced burning stages, are shown alongside regions where nucleosynthesis proceeds via neutron capture. The~iron (Fe) peak is marked to indicate the transition beyond which elements are predominantly produced through slow ($s$-process), intermediate ($i$-process), or~rapid ($r$-process) neutron capture.
Credit: S. Uthayakumaar, Facility for Rare Isotope Beams, Michigan State University, East Lansing, Michigan 48824, USA.  Nuclide chart (\url{https://people.physics.anu.edu.au/~ecs103/chart/}, accessed on 6 July 2025)   \hl{by} 
E. C. Simpson, Australian National University. This figure is reproduced with due~permission.}
\label{sivahami}
\end{figure}

Hydrogen, helium, and~a tiny amount of lithium are formed through nucleosynthesis in the Big Bang, but~their relative abundances are influenced by later processing in stellar interiors~\citep{bonifacio2007,masseron2012,bandyopadhyay4}. Beryllium and boron arise from spallation reactions in the interstellar medium. All other elements up to the iron peak are synthesized through nuclear fusion within stellar interiors or explosive nucleosynthesis in supernovae, as~shown in Figure~\ref{sivahami}.

In contrast, heavier elements (Z $>$ 30) are primarily formed through neutron-capture processes~\citep{frebelrev18,kobayashi2020}. The~rapid neutron-capture process ($r$-process) occurs in extreme conditions with high neutron flux, where the neutron-capture rate surpasses the beta-decay rate, leading to the formation of neutron-rich nuclei that subsequently decay to stability, producing the heaviest natural elements up to uranium. Approximately half of all stable isotopes of elements beyond the iron peak originate from this process, while the remaining half is produced through the slow neutron-capture process ($s$-process) or the intermediate neutron-capture process ($i$-process), both of which operate at different but lower neutron densities and~are thought to be primarily associated with asymptotic giant branch (AGB) stars. While the astrophysical sites of the $s$-process are relatively well understood and~progress is being made on understanding that of the $i$- process, the~dominant site of the $r$-process remains a topic of active debate, with~neutron star mergers being the only confirmed source. Other astrophysical sites of the $r$-process may also contribute, in~as-yet not understood~fractions.

Thus, the~formation of elements in the Universe is governed by a complex interplay of nucleosynthesis processes that span a wide range of astrophysical environments. A~fundamental challenge in astrophysics is to decipher the relative contributions of these processes to the chemical enrichment of the Milky Way and~to identify the astrophysical sites responsible for heavy-element production. Recent advancements in high-resolution spectroscopy have revolutionized our ability to probe the first generations of stars by enabling detailed elemental-abundance measurements, in~particular for the $r$-process-enhanced (RPE) stars identified by the $R$-Process Alliance (RPA)~\citep{christ2004,barklem2005,2siq2014,hansen18rpa,sakari18rpa,rana_rpa,rpa4,ban2024b}. These abundances serve as powerful tracers of the mass and explosion energies of the progenitor population(s), providing critical insights into the early initial mass function (IMF) and the chemical evolution of the Galaxy. {However, a~comprehensive understanding linking the nucleosynthesis of both light and heavy elements with the accretion history of the Galaxy, particularly across different stellar environments, in~order to trace the original sites of the $r$-process or early enrichment events and their relative contributions as a function of time, remains incomplete.}

As noted above, the~formation of $r$-process elements requires extreme conditions with high neutron fluxes, limiting their production sites to core-collapse supernovae, magneto-rotationally jet-driven supernovae, and~neutron star mergers (NSMs) or neutron star\hl{--}black hole mergers~\citep{thielemann2017}.\endnote{Recently, the~$i$-process has been suggested to also contribute a portion of the heaviest elements, particularly at the lowest metallicities (see~\citep{choplin2024} and~references therein).}

The recent detection of $r$-process elements in kilonovae following gravitational wave events~\citep{gw2017} has conclusively identified NSMs as at least one site of $r$-process nucleosynthesis~\citep{watson}. However, the~presence of RPE stars at very low metallicities suggests that heavy-element enrichment occurred early in the Universe. Such enrichments are likely to be preserved in environments with minimal subsequent star formation, such as in low-mass dwarf galaxies. Variations in $r$-process dilution levels within natal gas clouds likely influenced the observed dispersion in $r$-process enhancements, as~stars with higher [Fe/H] may have formed in higher-mass systems where the $r$-process yields were more diluted or where multiple progenitors contributed to~enrichment.

\begin{figure}[H]
\includegraphics[scale=0.55]{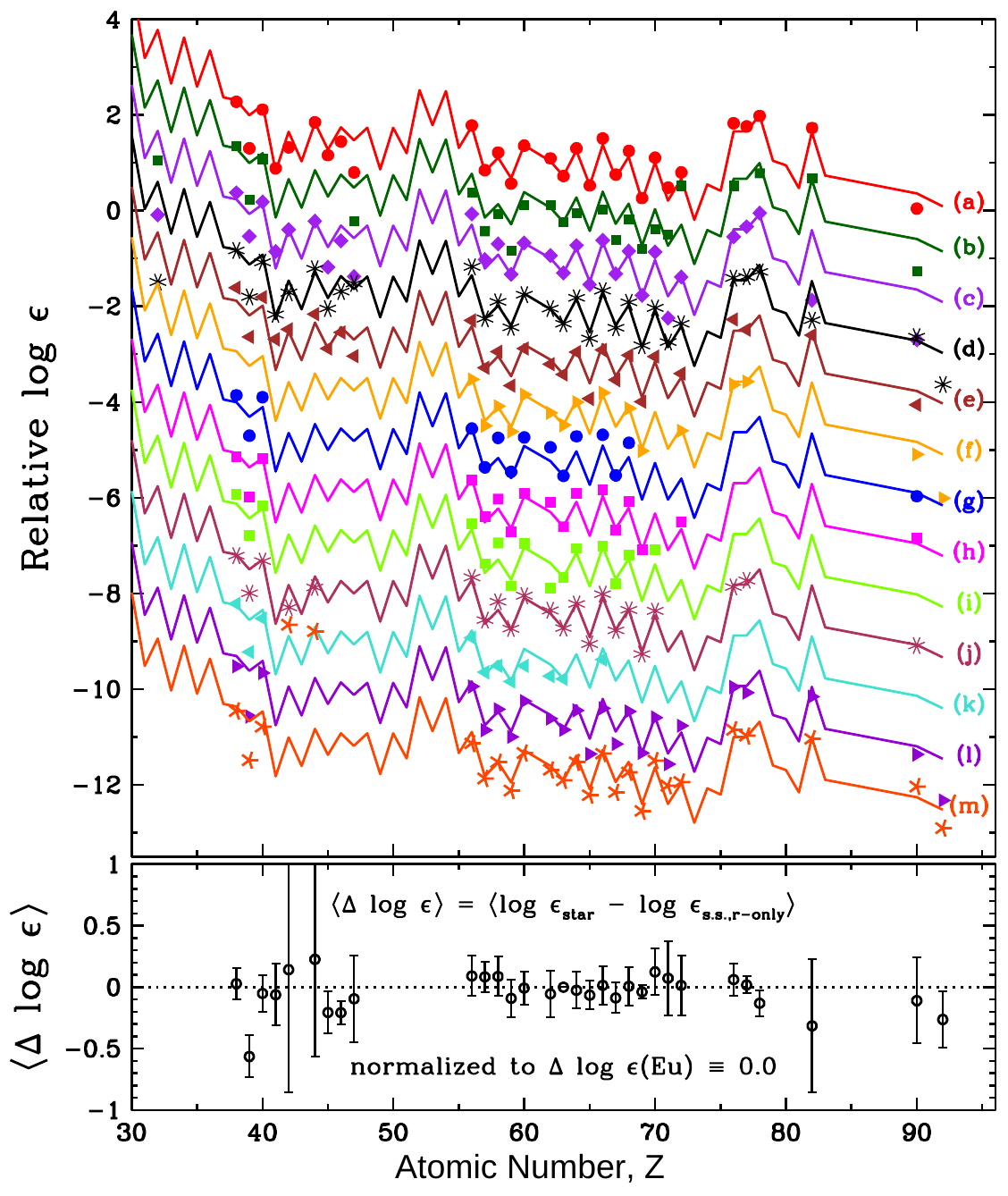}
\caption{\hl{This} 
figure, from~\citep{Cowan2021}, shows the robustness of the main $r$-process. Note that
log\,$\epsilon$(X) = log$_{10}$ \,($N_{\rm X}$/$N_{\rm H}) + 12$, where $N_\text{X}$ and $N_\text{H}$ represent the number densities of element X and hydrogen, respectively. Top panel: neutron-capture element abundances in 13 $r$-process-enhanced stars (individual data points), compared with the scaled Solar System $r$-process-only abundance pattern from~\citep{2siq2014}, primarily based on values from~\citep{simmerer2004}. All distributions have been normalized to match the europium abundance (Z = 63), with~additional vertical offsets applied to individual stellar patterns for clarity. Bottom panel: mean differences between the stellar abundances and the Solar System $r$-process pattern across the 13 stars. This figure is reproduced with permission and was originally published in \textit{Reviews of Modern Physics}.}
\label{cowan}
\end{figure}

\textls[-15]{{The robust universality of the main $r$-process pattern (near the second neutron-capture peak, containing elements like Eu, Ba, and others; see Figure~\ref{cowan}) across stars of all metallicities points to a dominant production mechanism (see~\citep{Cowan2021}).} However, the~large scatter in light $r$-process elements (first neutron-capture peak) and the presence of the so-called actinide-boost stars (stars that exhibit unusually high abundances of thorium and uranium relative to stable 2nd-peak elements such as europium) suggest multiple enrichment sites~\citep{holmbeck19,ban2020b,Wanajo2024}.} Observations of RPE ultra-faint dwarf (UFD) galaxies such as Reticulum II (Ret II;~\citep{jifrebelnature16,jifrebel2018,Ji2023}) further indicate that at least some of these sites contributed to rapid and early chemical enrichment, raising fundamental questions about the relative roles and timescales of different nucleosynthetic pathways in shaping the Milky Way’s chemical evolution.

To fully understand the cosmic origins of the astrophysical $r$-process, it is crucial to identify stars from partially or totally disrupted dwarf galaxies by taking advantage of precisely measured stellar dynamics and~employing them to trace their chemical evolution. The~enrichment levels of their $r$-process signatures coming from diverse sources (assuming more than one process contributes) will place valuable constraints on the role of NSMs in early epochs. Questions we need to address include whether NSMs were prolific across many progenitor satellite galaxies or~do we need additional astrophysical sites to explain the observed $r$-process signatures? {Furthermore, investigating the contributions from supernovae in these disrupted systems using the Fe-peak (elements in the periodic table around Fe such as Mn, Cr, Co, and Ni) and $\alpha$-elements (such as O, Mg, and Ca, which are produced by $\alpha$-capture reactions) will help determine the timescales of $r$-process enrichment and~whether multiple nucleosynthetic pathways were necessary to shape their chemical~evolution.}

\section{Signatures of the \emph{r}-Process in~Stars}

The $r$-process is (likely primarily) responsible for the formation of the heaviest elements in the Universe. This process unfolds in environments with extreme neutron densities, typically exceeding $n_n > 10^{22}$ cm$^{-3}$, where seed nuclei rapidly absorb neutrons. As~a result, highly neutron-rich isotopes form far from the valley of stability. Within~just a few seconds, these isotopes undergo a cascade of neutron captures before the neutron flux ceases, triggering radioactive decay that stabilizes them into heavy elements such as thorium (Z = 90) and uranium (Z = 92). Despite advances in nuclear physics and astrophysical modeling, predicting precise $r$-process abundance patterns remains a challenge. However, observational data, particularly from the Sun and ancient, metal-poor stars, reveals a distinct distribution of $r$-process elements~\citep{Sneden2008}.

\begin{figure}[H]

\includegraphics[scale=0.45]{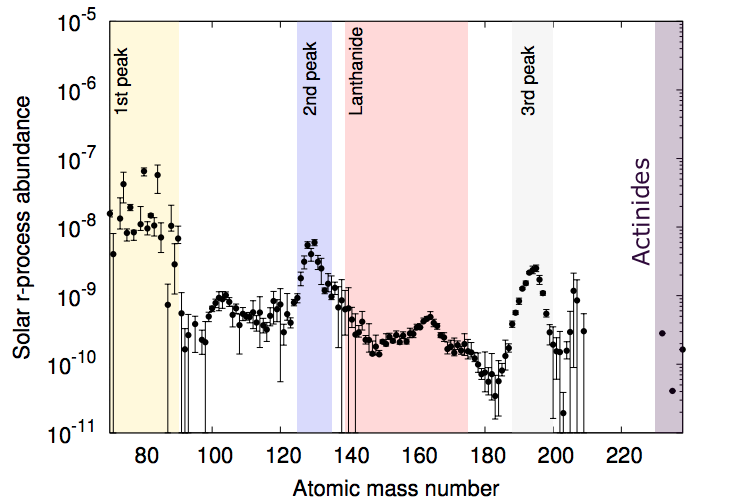}
\caption{\hl{The} 
three distinct $r$-process peaks as a function of the atomic mass number as observed in the Sun. The black filled circles represent the observed abundances corresponding to the atomic mass}
\label{cmd}
\end{figure}

A defining feature of the $r$-process is its characteristic abundance signatures, which consists of three major peaks in the relative abundances of elements (see Figure~\ref{cmd}). The~first peak, around Z = 34–40 (strontium, yttrium, and~zirconium, among~others), the~second near Z = 56–70 (barium, cesium, europium, and~others), and~the third around Z = 76–78 (e.g., osmium, iridium, and~platinum) arise due to the {delayed $\beta$ decay\endnote{Delayed $\beta$ decay is the process by which unstable, neutron-rich nuclei produced during the $r$-process undergo $\beta$ decay after neutron flux ceases, transforming into more stable elements and shaping the final abundance pattern.}} of neutron-rich progenitors with closed neutron shells. This mechanism fundamentally differs from the $s$-process, where neutron-magic nuclei accumulate at stable configurations, creating different peak~structures.

{\captionof*{figure} {number along with the uncertainties. This Figure was edited by E. M. Holmbeck, adapted from~\citep{Hotokezaka2018} with due permission from K. Hotokezaka, and was originally published in the \textit{International Journal of Modern Physics~D}.}}
\vspace{12pt}

For reference, Table~\ref{categories} summarizes the current definitions of the enrichment levels of stars, based on their iron-to-hydrogen ratios, relative to that of the \hl{Sun.}
\endnote{[Fe/H] = log$_{10}\,$($N$(Fe)/$N$(H))$_\star$ \hl{$-$} 
log$_{10}\,$($N$(Fe)/$N$(H))$_\odot$, where $N$(Fe) and $N$(H) represent the number densities of iron and hydrogen, respectively.
[X/Y] = log$_{10}\,$($N$(X)/$N$(Y))$_\star$ \hl{$-$} log$_{10}\,$($N$(X)/$N$(Y))$_\odot$, where $N$(X) and $N$(Y) represent the number densities of elements X and Y, respectively.
For sun, [Fe/H]$_\odot$ is defined as zero to set a reference point or benchmark.} {Note that, at~present, the~lowest metallicity stars have an evaluated upper limit (not detections) that approach [Fe/H] $= -8.0$, one hundred million times lower than the Solar level~\citep{keller2014}. Such upper limits are crucial to understand the enrichment processes and timescales of the early stellar populations and possibly the first stars.} The lowest metallicity RPE stars found to date occur at [Fe/H] $\sim -$3 to $\sim$ $-$4.0. None have yet been identified significantly below [Fe/H] = $-4.0$. This table also summarizes the different signatures of neutron-capture processes known at present and~the contemporary definition of carbon-enhanced metal-poor (CEMP) stars.

\begin{table}[H]
\tabcolsep2.5pt
\centering
\caption{\hl{Classes} 
and signatures of metal-poor~stars. \label{categories}}
\setlength{\tabcolsep}{-0.3mm}
\begin{tabular}{lcccc@{}}
\toprule
\textbf{Description} & \textbf{Definition} & \textbf{Abbreviation} \\
\midrule
Solar               & $\mbox{[Fe/H]} =0.0$  &      \\
Metal-poor          & $\mbox{[Fe/H]}\leq-1.0$  &   MP \\
Very metal-poor     & $\mbox{[Fe/H]}\leq-2.0$  &   VMP \\
Extremely metal-poor& $\mbox{[Fe/H]}\leq-3.0$  &   EMP \\
Ultra metal-poor    & $\mbox{[Fe/H]}\leq-4.0$  &   UMP \\
Hyper metal-poor    & $\mbox{[Fe/H]}\leq-5.0$  &   HMP \\
Mega  metal-poor    & $\mbox{[Fe/H]}\leq-6.0$  &   MMP \\
Septa metal-poor    & $\mbox{[Fe/H]}\leq-7.0$  &   SMP \\
Octa metal-poor     & $\mbox{[Fe/H]}\leq-8.0$  &   OMP \\
Giga metal-poor     & $\mbox{[Fe/H]}\leq-9.0$  &   GMP \\
\midrule

\textbf{\hl{Signature}
}               & \textbf{\hl{Criteria}}    &\textbf{\hl{Abbreviation}}\\\midrule

\multirow{2}{*}{\hl{$r$-process-enhanced}
}       &  $+0.3 < \mbox{[Eu/Fe]} \le +0.70$ and $\mbox{[Ba/Eu]} < 0.0$ & $r$-I  \\
&  $\mbox{[Eu/Fe]} > +0.7$ and $\mbox{[Ba/Eu]} < 0.0$           & $r$-II \\\midrule

limited $r$-process     & $\mbox{[Eu/Fe]} \leq +0.3$, $\mbox{[Sr/Ba]} > +0.5$, and~$\mbox{[Sr/Eu]} > 0.0$ & $r_{lim}$ \\\midrule

$s$-process:            &  $\mbox{[Ba/Fe]} > +1.0$, $\mbox{[Ba/Eu]} > +0.5$; also $\mbox{[Ba/Pb]} > -1.5$        & $s$\\\midrule

$r$- and $s$-process    & $0.0 < \mbox{[Ba/Eu]} < +0.5$ and $-1.0<\mbox{[Ba/Pb]}<-0.5$ &$r/s$\\\midrule

$i$-process             &   $0.0 < \mbox{[La/Eu]} < +0.6$ and $\mbox{[Hf/Ir]}\sim +1.0$ & $i$\\\midrule

carbon-enhanced       & $\mbox{[C/Fe]} > +0.7$ &  CEMP\\
\bottomrule

\end{tabular}
\end{table}

The presence of $r$-process elements in extremely metal-poor ([Fe/H] $\leq -3.0$) stars confirms that this process was active in the early Universe. These rare stars serve as fossil records of ancient nucleosynthesis, preserving the chemical imprints of past $r$-process events. Large-scale surveys suggest that only 3–5\% of stars with [Fe/H] $< -2.5$ in the Galactic halo system exhibit strong $r$-process enhancement~\citep{christlieb2003, barklem2005, hansen18rpa, sakari18rpa, rana_rpa, rpa4, ban2024b}. These stars did not synthesize neutron-capture elements themselves; rather, they inherited them from an earlier generation of stellar explosions. The~striking similarity between the main $r$-process patterns in these stars and the Solar $r$-process signature suggests a common nucleosynthetic~origin.

The degree of $r$-process enrichment in metal-poor stars is commonly measured using the europium-to-iron ratio relative to the Sun, [Eu/Fe], which serves as a proxy for neutron-capture element production~\citep{beers2005}. Based on this metric, RPE stars are classified into two primary categories: $r$-I stars, with~moderate enhancement (\mbox{+0.3 $<$ [Eu/Fe] $\leq$ +0.7)} and~$r$-II stars, which exhibit strong enhancement ([Eu/Fe] $>$ +0.7). The~latter group, with~europium levels more than five times the Solar value, provides compelling evidence for localized $r$-process events. Some of the most well-known $r$-II stars include \mbox{CS 22892-052}~\citep{sneden2000}, the~first identified $r$-II star with [Fe/H] $\lesssim -3.0$. This star is also enhanced in carbon (and hence can be classified as CEMP-$r$, following the nomenclature of~\citep{beers2005}). Others include CS 31082-001~\citep{cayrel2001}, which is not carbon enhanced but~exhibits thorium, uranium, and~an actinide boost (described below), and~HE 1523-0901~\citep{Frebel2007}, which also contains uranium. These stars, despite their extremely low metallicities, exhibit an enrichment of neutron-capture elements by factors of 40–70 relative to iron. Note also that there are a few metal-poor stars with enrichment levels exceeding 100 (e.g., 2MASS J15213995-3538094~\citep{Cain2020} and~2MASS J22132050–5137385~\citep{Roe2024}), which are sometimes referred to as $r$-III (\mbox{[Eu/Fe] $> +2.0$}) stars~\citep{hirai2024}; this definition may shift as more such stars are~identified.

\begin{figure}[H]
\hspace{-6pt}
\includegraphics[scale=0.75]{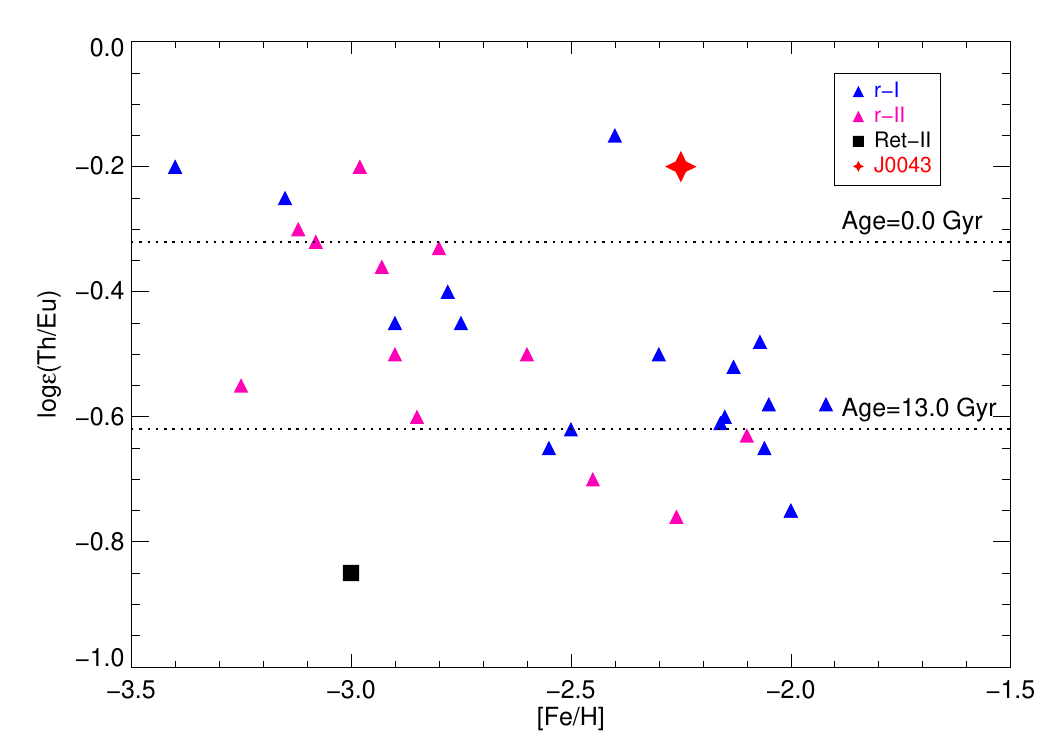}
\caption{\hl{Distribution} 
of log$\,\epsilon$(Th/Eu) for metal-poor stars as a function of [Fe/H]. Previously recognized $r$-I (blue triangles) and $r$-II (pink triangles) stars are shown for comparison, along with a bright $r$-II star (black square) from a dwarf galaxy. Dashed lines indicate corresponding ages derived from the [Th/Eu] ratios. Stars with unusually high Th relative to Eu, known as actinide-boost stars, occupy the upper region of the diagram and~yield un-physically young ages due to their elevated [Th/Eu] values. Figure adapted from~\citep{ban2020b}, reproduced with permission © AAS.}
\label{act-boost}
\end{figure}

The chemical-enrichment history of these stars rules out localized atmospheric effects as the cause of their $r$-process signatures. Radial velocity monitoring studies (e.g.,~\citep{Hansen2015} and~references therein) reveal that approximately 80\% of RPE stars exhibit no evidence of binary motion, eliminating mass transfer from a companion as a likely explanation. Instead, their neutron-capture enrichment reflects the properties of the natal gas clouds from which they formed, shaped by prior supernovae or NSMs. However, the~existence of actinide-boost stars (e.g.,~\citep{hill2002,Mashonkina2014,Holmbeck2018,jifrebel2018,Placco2023}) complicates this picture, hinting at possible variations in $r$-process yields or multiple contributing astrophysical sources.
See Figure~\ref{act-boost}.

\vspace{-6pt}

Recent observational evidence suggests that fission fragments from transuranic nuclei may significantly influence the observed abundance patterns of RPE stars. \hl{Ref.}
~\citep{Roe2023sci} analyzed a sample of 42 RPE stars and identified strong correlations between the abundances of certain lighter (Z = 44–47) and heavier (Z = 63–78) $r$-process elements. These correlations, not seen for adjacent atomic numbers, are inconsistent with traditional two-component $r$-process models, and~are instead interpreted as signatures of fission products originating from very neutron-rich nuclei with mass numbers exceeding 260. The~presence of nearly constant abundance ratios, such as [Ag/Eu], across the stellar sample further supports the role of fission in smoothing out variations introduced by initial conditions at the nucleosynthesis site. This highlights fission as a potentially universal feature in $r$-process events, such as neutron star mergers, contributing to the chemical makeup of the early~Galaxy.

\textls[-25]{The broader population of RPE stars spans a metallicity range of $-4.0 <$ [Fe/H] $< -1.5$. However, at~a metallicity of $\sim -1.5$, the~median [Eu/Fe] among Galactic halo stars drops to approximately +0.3~\citep{frebelrev18}. At~this point, nearly 50\% of stars satisfy the $r$-I classification, making the definition of RPE stars less distinct. To~confirm whether these stars genuinely reflect an $r$-process enrichment history, one must examine their complete neutron-capture element abundance patterns. The~influence of other nucleosynthetic pathways, such as from the $s$-process, becomes evident at higher~metallicities.}

Despite considerable progress, the~identification of $r$-process production sites remains an open question. New observational and theoretical developments continue to refine our understanding of how and where these elements are synthesized. The~next sections explore the most promising astrophysical candidates, focusing on core-collapse supernovae, magneto-rotational explosions, and~neutron star mergers, and~their respective contributions to Galactic chemical~evolution.

\section{Astrophysical Sites for the Production of \emph{r}-Process~Elements} \label{sec:floats}

The fundamental question in $r$-process nucleosynthesis is identifying the astrophysical sites responsible for producing heavy elements via rapid neutron capture~\citep{th2011,tsuji1,Schatz2022,banchemo}. The~primary requirement for an $r$-process site is an environment with an abundant neutron supply, ensuring a sufficiently high neutron-to-seed ratio for the synthesis of heavy nuclei~\citep{frebelrev18}. Several astrophysical events have been proposed as potential $r$-process sites, each contributing to different phases of cosmic~evolution.

\begin{figure}[H]

\includegraphics[scale=1.2]{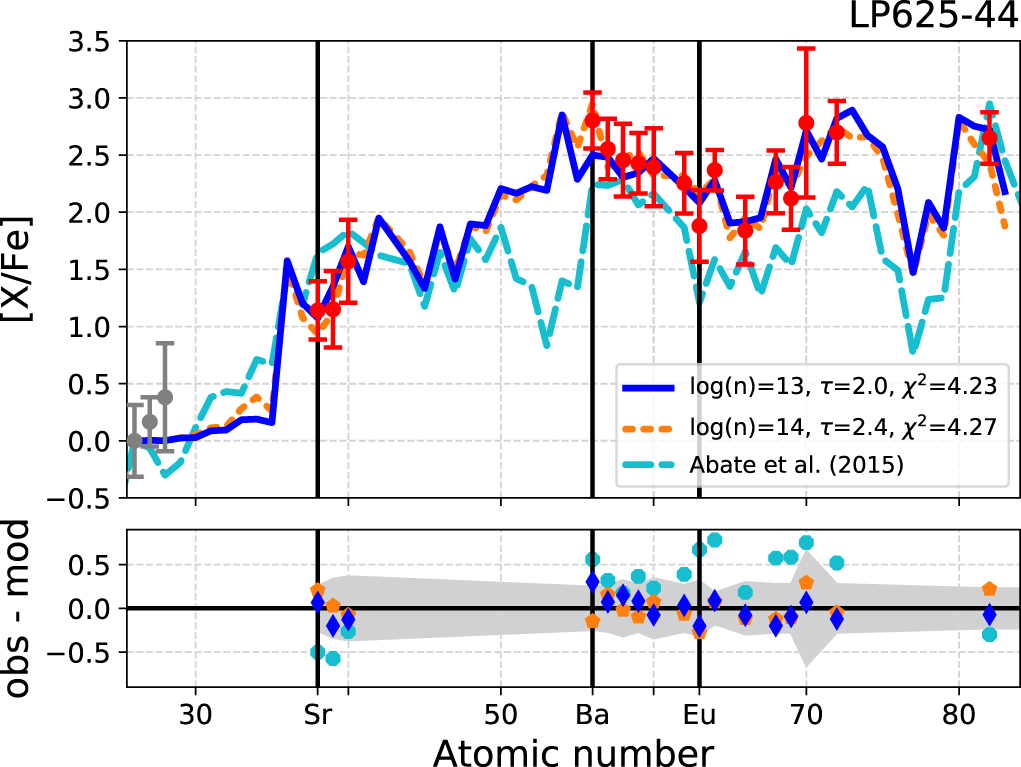}
\caption{\hl{Best} 
fits to the heavy-element abundance pattern of the CEMP-$i$ star LP625-44.  In~this figure, taken from~\citep{Hampel2019}, the~blue and orange lines represent the best-fitting $i$-process models with neutron densities of $n = 10^{15}~\mathrm{cm}^{-3}$ and $n = 10^{14}~\mathrm{cm}^{-3}$, respectively. The~lower panel shows the residuals of the measured abundances from the adopted $i$-process model, with~the expected range of uncertainties in the model indicated in shaded gray. This figure is reproduced with permission © AAS and was originally published in \textit{The Astrophysical~Journal}.}
\label{iproc-hampel}
\end{figure}

One of the key physical parameters used to distinguish $r$-process sites is the neutron-to-seed ratio, usually quantified by the electron fraction (Y$_{\rm e}$). Higher neutron-to-seed ratios lead to lower Y$_{\rm e}$ and vice~versa. A~low electron fraction (Y$_{\rm e} < $ 0.10), corresponding to high neutron densities, results in the formation of the heaviest nuclei, including the third $r$-process peak (e.g., actinides such as thorium and uranium). In~contrast, environments with moderate electron fractions predominantly produce elements associated with the main $r$-process, while higher electron fractions (Y$_{\rm e} > 0.25$) lead to the synthesis of only the first $r$-process peak elements. However, additional processes and mechanisms, such as the $s$-process and the $i$-process, can also contribute to elements near the first peak (and beyond for~the $i$-process; see Figure~\ref{iproc-hampel}), necessitating a careful assessment of their~origins.

Among the most widely considered astrophysical sites for $r$-process nucleosynthesis are core-collapse supernovae and binary neutron star mergers, along with other potential sources such as such as collapsars (massive star collapses leading to black holes with accretion disks) and~neutron star\hl{--}black hole mergers, which may also contribute to $r$-process elements under specific~conditions.

\subsection{Neutron Star~Mergers}

NSMs are widely recognized as a key site for $r$-process nucleosynthesis, where rapid neutron captures form the heaviest elements in the Universe. These mergers occur when two compact objects, such as neutron stars or a neutron star and a black hole, gradually spiral inwards while losing energy through gravitational wave emission. The~groundbreaking gravitational wave detection GW170817~\citep{gw2017}, accompanied by the kilonova AT2017gfo~\citep{drout2017,perego2017}, provided definitive observational evidence that binary neutron star (BNS) mergers can produce heavy $r$-process elements. The~detection of strontium in the electromagnetic counterpart confirmed the synthesis of lighter $r$-process nuclei~\citep{watson}, while the broad-band light curves~\citep{drout2017} and spectral evolution~\citep{shappe2017} indicated the presence of both light and heavy $r$-process components. A~summary of the multi-messenger results from this event is provided by~\citep{gw2017}.

BNS mergers are expected to eject roughly 0.01 M$_\odot$ of neutron-rich material during the collision, with~the precise ratio of light-to-heavy $r$-process elements varying based on the ejecta's neutron fraction and the mass and spin of the compact objects (e.g.,~\citep{Holmbeck2024} and~references therein). Simulations have shown that BNS mergers can generate significant amounts of both the first and second $r$-process peak elements, while neutron star\hl{--}black hole (NSBH) mergers could play a more substantial role in forming the heaviest $r$-process species~\citep{Siegel2017,Wehmeyer2019}, depending on the black hole's spin and the tidal disruption dynamics~\citep{Roberts2017}. However, the~overall contribution of NSMs to the Galactic $r$-process inventory remains debated, as~their suggested delayed timescales and potential for substantial velocity kicks may cause enrichment to occur long after or far away from the formation of many~stars.

The delay time between the formation of the progenitor stars and the actual merger event is governed by the timescale required for the system to lose energy through gravitational radiation. Population-synthesis models suggest that most NSMs occur with a delay of tens to hundreds of millions of years, although~a small fraction may merge rapidly, within~a few Myr~\citep{Ishimaru2015,Maoz2025}. These variations complicate the role of NSMs in early Galactic chemical evolution, as~mergers with longer delay times may not contribute to the $r$-process enrichment of low-metallicity stars. Recent studies indicate that NSBH mergers, with~their shorter in-spiral timescales and higher ejecta masses, may have played a more significant role in enriching the early Milky Way with $r$-process elements~\citep{Wehmeyer2019,Koba2023}.

As noted above, despite the clear evidence that NSMs produce $r$-process material, their overall contribution to the Galactic $r$-process inventory is still uncertain. Some chemical-evolution models, such as those \hl{by}
~\citep{Hotokezaka2018,kobayashi2020}, suggest that NSMs alone cannot account for the entire $r$-process abundance distribution, particularly at low metallicities. These models indicate that mergers would need to occur with unusually short delay times or exceptionally high ejecta masses to explain the observed europium distribution in metal-poor stars. Alternatively, an~additional astrophysical site, such as rare magneto-rotational supernovae or neutron star\hl{--}white dwarf (NSWD) mergers, may be required to fully account for the Galactic $r$-process~enrichment.

Recent observations of kilonova-like transients following long-duration gamma-ray bursts (GRBs), such as GRB 211211A and GRB 230307A, have sparked interest in NSWD mergers as potential $r$-process sites~\citep{Levan}. However, current estimates suggest that NSWD mergers eject significantly less neutron-rich material than NSMs or NSBH mergers, likely contributing less than 0.0005 M$_\odot$ of $r$-process material per event, making them an unlikely dominant source~\citep{Chen2024}.

Three-dimensional magneto-hydrodynamic simulations of NSM remnants
have revealed that the neutron-rich ejecta exhibit a broad range of Y$_e$, spanning from $\sim$0.25 to $0.40$. This diversity leads to an $r$-process abundance pattern that does not perfectly match the Solar pattern, as~the production of the heaviest $r$-process species remains limited. \hl{}
Ref.~\citep{Holmbeck2024} further demonstrated that the present-day Galactic population of BNS systems tends to favor the production of lighter $r$-process elements, which could imply the need for additional sites to explain the Solar System's abundance of the heaviest $r$-process~species.

Overall, while NSMs are confirmed as prolific $r$-process factories, ongoing debate persists regarding their relative importance. Their suggested delayed timescales, potential for high-velocity kicks, and~the challenges in reconciling their yields with observed stellar abundances indicate that complementary astrophysical sites may still be required to fully explain the origin of the heaviest elements in the Milky~Way.

\subsection{Core-Collapse~Supernovae}

The idea that core-collapse supernovae (CC-SNe) are potential sites for $r$-process nucleosynthesis has been explored since the pioneering work of~\citep{burbidge1957,Cameron1957}. However, subsequent observations and models (e.g.,~\citep{mcwilliam1998}) have shown that not all supernovae are equally capable of producing $r$-process elements. Instead, only a small fraction, estimated at around 1–10\%, of~core-collapse events may generate the conditions necessary for synthesizing heavy $r$-process nuclei~\citep{nishimura2015,Shibagaki2016,Nishimura2017}, including those around the second and third abundance peaks. This implies that $r$-process nucleosynthesis in CC-SNe is relatively rare but~potentially plays a significant role in Galactic chemical~evolution.

Several mechanisms have been proposed to explain how CC-SNe might produce the full range of $r$-process elements. One of the most studied scenarios involves neutrino-driven winds emanating from the proto neutron star~\citep{Nevin2023}. As~the star collapses, intense neutrino fluxes expel neutron-rich material, creating conditions suitable for rapid neutron capture. Electron-capture supernovae, originating from progenitors with masses between roughly 8 and 10 \hl{M}
$_\odot$, have also been considered as potential $r$-process sites~\citep{Jones2019}. However, modern hydrodynamical simulations indicate that these mechanisms often fail to generate the neutron richness required for a complete $r$-process, instead producing a weaker $r$-process that is limited to lighter neutron-capture elements near the first $r$-process peak~\citep{Jones2016,Farouqi2022,Zha2022}. This limited $r$-process is believed to occur more frequently, and~may well account for the light neutron-capture elements such as Sr, Y, and Zr observed in some metal-poor~stars.

From an observational perspective, metal-poor stars that exhibit signatures of a limited $r$-process provide key insights~\citep{snedenparthasarathy1983,honda2006,Sylakis2024} on the possible progenitors such as CC-SNe~\citep{Casey2017}. These stars, now classified as ``limited $r$-process stars,'' exhibit abundance patterns characterized by [Eu/Fe] $\leq$ +0.3, [Sr/Ba] $>$ +0.5, and~[Sr/Eu] $>$ 0.0~\citep{rpa4}. The~chemical signatures of these stars suggest that they formed from gas enriched by weak $r$-process events, likely from CC-SNe. Theoretical models propose that this limited $r$-process originates from a quasi-statistical equilibrium (QSE) phase with a relatively low neutron-to-seed ratio ($<$100), which prevents the synthesis of heavier $r$-process elements. Consequently, only lighter neutron-capture elements near the first peak are produced, with~rapidly diminishing yields toward the second peak~\citep{Sylakis2024}.

{The weak $r$-process also operates on even shorter timescales compared to the main $r$-process. In~such environments, often associated with neutrino-driven winds or early core-collapse 
supernovae, ($\alpha$,n) reactions can play a critical role by enabling the material to reach higher atomic numbers, effectively bridging gaps that would otherwise require slow $\beta$-decays. This pathway has attracted considerable attention, both theoretically and experimentally, with~recent work focusing on reaction rates and nuclear inputs relevant to light trans-iron nuclei. These developments help refine predictions for the light $r$-process element patterns observed in many metal-poor stars. For~more details, see~\citep{arcones2011,Mum2020,Wang2024}.}

Recent studies using magneto-hydrodynamical simulations have added new insights into $r$-process production in CC-SNe. Models by~\citep{nishimura2015} explore the role of rotation and magnetic fields in influencing the nucleosynthesis outcomes. Their work identifies two explosion scenarios: prompt magnetic-jet and delayed magnetic-jet explosions. In~the prompt magnetic-jet case, where the magnetic fields are particularly strong, heavy $r$-process elements, including actinides, are synthesized. In~contrast, the~delayed magnetic-jet explosions, associated with weaker magnetic fields, tend to produce only lighter $r$-process nuclei, up~to the second $r$-process peak (A $\sim$ 130). These findings suggest that the strength of the magnetic fields in CC-SNe significantly influences the extent of $r$-process nucleosynthesis, with~only the most magnetically energetic events contributing to the production of the heaviest $r$-process~elements.

Contemporary models suggest that the most promising CC-SNe sites for full $r$-process nucleosynthesis involve extreme physical conditions, such as those found in magneto-rotationally driven jets~\citep{nishimura2015,Reichert2022,Zha2024} and collapsar disk winds~\citep{,Miller2020,Barnes2022}. In~these rare events, extremely rapid rotation and strong magnetic fields create the necessary neutron-rich ejecta for synthesizing heavy $r$-process nuclei. These models also indicate that such rare supernovae, if~they occur, could generate substantial $r$-process yields (0.01–0.1 M$_\odot$) immediately following episodes of star formation, given the short lifetimes of their progenitors. Interestingly, collapsars have also been argued as prodigious producers of carbon, and~may be responsible for the origin of the CEMP-$r$ stars~\citep{siegel2019}.

\begin{figure}[H]

\includegraphics[scale=0.5]{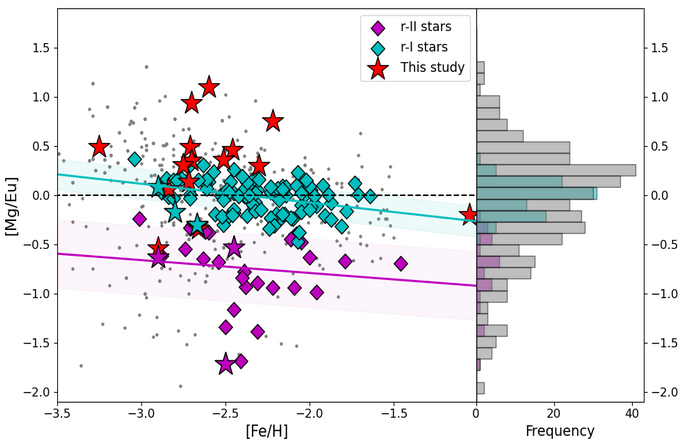}
\caption{\hl{Distribution} 
of [Mg/Eu], as~a function of [Fe/H], for $r$-I, $r$-II, and~non-$r$-process-enhanced (non-RPE) stars. Red stars indicate non-RPE stars, while $r$-I and $r$-II stars are taken from RPA Data Releases 1--5, with~colors assigned as in the legend. Gray-filled circles represent stars from~\citep{jinabase}. The~histogram on the right compares the [Mg/Eu] distributions across all metallicities. The~observed downward trend of [Mg/Eu] with increasing [Fe/H] suggests the growing importance of a delayed nucleosynthetic source, distinct from CC-SNe that preferentially contributes europium as the Universe evolves. Figure credits: \cite{ban2024b}, reproduced with permission © AAS and was originally published in The Astrophysical Journal Supplement~Series.}
\label{rpa5}
\end{figure}

The interplay between the limited and main $r$-processes offers a compelling explanation for the observed abundance patterns in RPE metal-poor stars. If~the main $r$-process produces second- and third-peak elements, while CC-SNe generate variable amounts of lighter $r$-process material, the~resulting abundance patterns in metal-poor stars could reflect a mixture of these different nucleosynthetic processes. The~presence of small, yet detectable, amounts of neutron-capture elements in most metal-poor stars could be attributed to ongoing chemical enrichment by CC-SNe, which contribute both light and heavy $r$-process material over time. Recent results from the RPA DR5 reveal that the [Mg/Eu] ratio in RPE stars, for~both $r$-I and $r$-II classes, systematically decreases with increasing metallicity, as~shown in Figure~\ref{rpa5}. This trend suggests that europium, primarily produced by the $r$-process, becomes increasingly abundant relative to magnesium, which is synthesized in CC-SNe. The~observed behavior implies the emergence or growing contribution of additional $r$-process sources distinct from core-collapse supernovae as the Universe evolves and becomes more~metal-rich.

Interestingly, stars such as HD 122563~\citep{honda2006} exhibit abundance patterns that may offer a rare glimpse of a
``pure'' limited $r$-process signature. This metal-poor star shows a pronounced over-abundance of light neutron-capture elements relative to heavier ones, consistent with enrichment by CC-SNe ejecta that produced only first-peak $r$-process elements. However, the~contribution of charged-particle reactions and $s$-process nucleosynthesis in massive stars may also play a role in shaping the abundance patterns of certain early-generation stars~\citep{Cescutti2013,Prantzos2018}.

Overall, while CC-SNe are no longer considered the dominant site for the heaviest $r$-process elements~\citep{FrebelJi2023,banchemo}, they remain important contributors to the chemical enrichment of the Galaxy, especially through their production of lighter $r$-process nuclei. The~complexity of the nucleosynthesis processes, combined with the rarity of extreme explosion conditions, makes the $r$-process in CC-SNe an ongoing area of active research, with~implications for the chemical evolution of galaxies and the formation of metal-poor~stars.

\subsection{Other~Sites}

As mentioned above, the~$i$-process has been proposed as another source of heavy elements beyond the iron peak, up~to and including the lanthanides and actinides (\citep{Hampel2019,choplin2024} and~references therein). Originally suggested by~\citep{cowanandrose}, the~discovery of CEMP stars with [C/Fe] $>$ +0.7 and [Fe/H] $\leq -1.0$ exhibiting over-abundances of \hl{both} 
$r$- and $s$-process elements (CEMP-$r/s$; see~\citep{beers2005}) inspired models to account for this remarkable behavior. The~likely astrophysical sites for the $i$-process include low-mass, low-metallicity AGB stars~\citep{choplin2024} and rapidly accreting white dwarfs (\citep{Den2019} and~references therein). Only a few tens of CEMP-$r/s$ stars are known at present, but~among those that are known, the~binary fractions appear high, at~least $\sim$50\%~\citep{jinmiyoon}. Definitive assignment of the astrophysical site(s) of the $i$-process awaits the discovery and high-resolution spectroscopic analysis of additional CEMP-$r/s$ stars.

\section{Astrophysical Environments for Studying the \emph{r}-Process}
\unskip

\subsection{Dwarf~Galaxies}

The study of $r$-process nucleosynthesis in dwarf galaxies presents unique challenges, largely due to the intrinsic faintness of their stellar populations. Dwarf galaxies, loosely defined as gravitationally bound systems of stars embedded in dark matter halos, are significantly farther away, and~hence much fainter than nearby Galactic stars~\citep{McConnachie2012,Simon2019,McConnachie2020}. Thus, their detailed spectroscopic analysis is technically demanding. High-resolution spectroscopy of individual stars in these systems requires the use of large-aperture telescopes and long integration times, often pushing the limits of current instrumentation. Furthermore, observations in the near-UV, where several key neutron-capture element lines are located, remain infeasible for such distant and faint targets with present-day ground-based~facilities.

Despite the above difficulties, dwarf galaxies offer a valuable window into the early Universe. Their relatively simpler and truncated star-formation histories preserve the chemical signatures of early nucleosynthesis events more clearly than for individual stars in the more chemically evolved Milky Way halo system. As~such, they provide a unique opportunity to probe the conditions and sites responsible for the origin of heavy elements. Measurements of neutron-capture elements like strontium, barium, and~europium in individual red giant stars, among~the only accessible neutron-capture tracers in these systems, allow us to constrain the yields and delay times of the astrophysical events responsible for $r$-process enrichment. To~date, roughly 60 dwarf galaxies are known to orbit the Milky Way~\citep{Drlica2020}, and~they continue to serve as prime laboratories for investigating the nature of early chemical enrichment and the role of $r$-process nucleosynthesis in the broader context of galaxy~evolution.

\begin{figure}[H]\hspace{-6pt}
\includegraphics[scale=0.22]{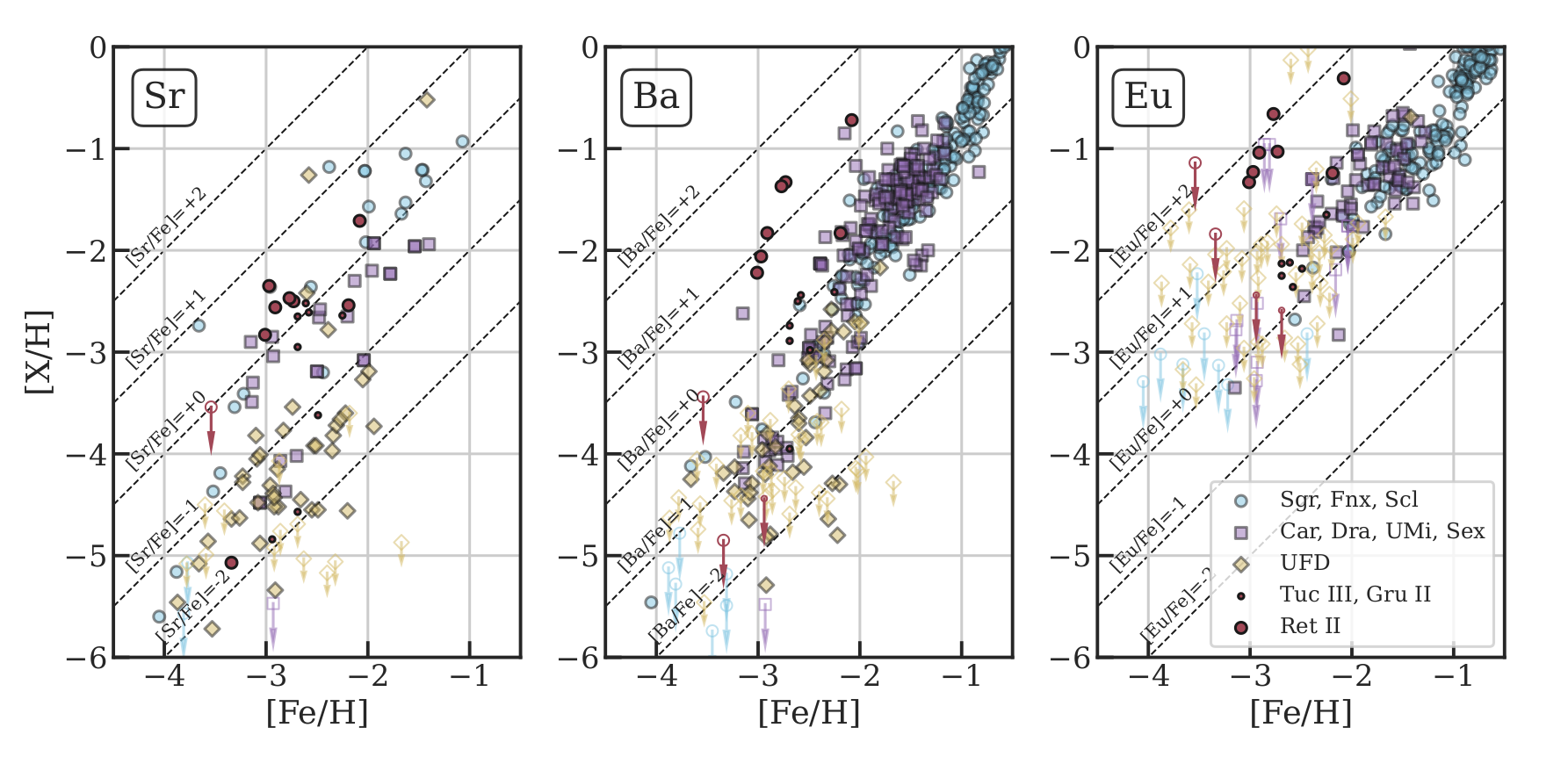}
\caption{\hl{Neutron}
-capture element abundances in dwarf galaxies vary with stellar mass. In~this figure, taken from~\citep{FrebelJi2023}, blue circles represent more massive classical dSphs, purple squares correspond to intermediate-mass classical dSphs, yellow diamonds mark UFDs without $r$-process signatures, and~red circles highlight the few UFDs with $r$-process enhancement. Reticulum II (Ret II) is shown with larger symbols to emphasize its higher [Eu/Fe] compared to Tucana III and Grus II. Open symbols with downward arrows indicate upper limits where no detection was possible. Most
UFDs have lower neutron-capture abundances than classical dSphs across $-3 < \mathrm{[Fe/H]} < -2$, except~for the rare $r$-process-enhanced systems. Upper limits for strontium and barium are often meaningful, while those for europium mainly reflect the difficulty of its detection at low abundances. The~increase in barium at $\mathrm{[Fe/H]} \geq -2$ among classical dSphs and some UFDs indicates the onset of significant $s$-process enrichment. The~figure is reproduced with permission from the authors.}
\label{dwarfgal}
\end{figure}

A particularly compelling case for $r$-process enrichment in dwarf galaxies emerged with the discovery of the UFD galaxy Ret II (see Figure~\ref{dwarfgal}). High-resolution spectroscopic observations~\citep{jifrebelnature16,jifrebel2018,Ji2023} have revealed that the majority of its red giant stars exhibit extreme enhancements in $r$-process elements, some with europium abundance ratios exceeding [Eu/Fe] $>$ +1.5 and~clear signatures matching the Solar $r$-process pattern. This finding provided the first strong evidence that a single, prolific $r$-process event such as a NSM could enrich an entire dwarf galaxy in the heaviest elements. The~sharp chemical imprint and the lack of a large dispersion in the $r$-process abundance patterns among the RPE stars suggest that the $r$-process material was injected into the interstellar medium rapidly and uniformly, before~any significant dilution or subsequent star formation~occurred.

RPE stars have also been found in other dwarfs, such as Tucana III~\citep{Marshall2019} and possibly in the more massive Sculptor and Fornax dwarf galaxies~\citep{Reichert2021,Lucchesi2024}, although~with more complex enrichment histories. These discoveries support the idea that NSMs, despite their presumed long delay times, can occur early and contribute significantly to $r$-process enrichment even in the smallest galactic environments. They also offer critical insight into the stochastic nature of nucleosynthetic events in low-mass systems, highlighting how a single rare event can dominate the chemical evolution of an entire~galaxy.

The variation in $r$-process enrichment across dwarf galaxies provides crucial constraints on the frequency and environmental dependence of $r$-process events. While systems like Ret II display extreme enhancements, many other dwarf galaxies, particularly the classical and~more massive ones, exhibit only mild or negligible levels of $r$-process element enrichment. For~instance, in~dwarf galaxies like Draco or Ursa Minor, only a small fraction of stars exhibit detectable europium, and~the abundances often reflect more modest enrichment, possibly from multiple weaker sources or events with lower~yields.

\vspace{-6pt}

The wide range of enrichment levels suggests that $r$-process production in dwarf galaxies is not uniform but instead stochastic, likely depending on factors such as star-formation efficiency, gas retention, and~the timing and/or location of explosive events. Ultra-faint dwarfs, with~their limited star formation and shallow potential wells, are especially susceptible to enrichment from a single nucleosynthesis event, making them ideal laboratories to study the imprint of individual $r$-process progenitors. In~contrast, more massive dwarfs may host a more prolonged and mixed chemical evolution, where the signal from $r$-process events becomes diluted or confounded with contributions from other processes. By~comparing the chemical patterns across systems with different masses and star-formation histories, astronomers can begin to piece together a more complete picture of how and when the $r$-process shaped the early chemical evolution of galaxies~\citep{Hirai2025}.

\subsection{Globular~Clusters}

\textls[-15]{Globular clusters (GCs), once regarded as simple, chemically homogeneous stellar populations that were formed in a single burst, have undergone a dramatic conceptual transformation. This shift began in the 1970s with the groundbreaking discovery of star-to-star abundance variations in light elements like carbon and nitrogen~\citep{kraftb}, followed by sodium and oxygen~\citep{peterson1980,kraft1997}. These early findings changed the classical paradigm, revealing that GCs harbor multiple stellar populations with distinct chemical fingerprints~\citep{gratton2004,gratton2012,gratton2019}.} Spectroscopic studies, including the extensive survey by~\citep{carretta2009a,carretta2009b}, solidified this picture, demonstrating that light-element anti-correlations (such as Na-O) are a defining feature of nearly all Galactic GCs. Complementing this, high-precision HST photometry
\citep{piotto2009,milone2017,milone2019,milone2022} uncovered distinct photometric sequences corresponding to chemically distinct sub-populations, linking spectroscopic abundance patterns with variations in He, C, N, and~O. While variations in light elements could be traced to high-temperature hydrogen burning in a previous generation of stars, the~detection of inhomogeneities in heavier elements, especially those forged in the $r$-process, presented a far more profound and intriguing~challenge.

The first compelling evidence for such $r$-process dispersion came from M15. 
Ref.~\citep{Sneden1997m15} discovered strong star-to-star variations in Eu and Ba, with~a remarkably constant [Eu/Ba] ratio pointing to a pure $r$-process origin. This suggested that a singular, potent nucleosynthetic event may have seeded the cluster with both of these elements. Follow-up studies~\citep{sneden2000m15,otsuki2006,sobeck2011m15,Worley2013} confirmed this picture and, crucially, demonstrated that these heavy-element anomalies were decoupled from the canonical light-element trends. In~M15, stars enriched in Na and depleted in Mg bore no connection to their $r$-process content, pointing to distinct enrichment~pathways.

Explaining these abundance patterns turned out to be challenging. Theories invoking neutron star mergers (NSMs) post-star formation~\citep{tsuji1,tsuji2} were undermined by observations like those of~\citep{kirby2020}, who showed constant [Ba/Fe] across evolutionary stages\hl{--}contradicting expectations from surface pollution. More plausible are models invoking rare, early in situ events such as magneto-rotational supernovae or NSMs that enriched the proto-cluster gas before it fully mixed~\citep{Tarumi2021}, leaving behind localized chemical signatures in the first generation of stars. For~a time, M15 stood as a lone anomaly. Ref. \citep{RoeSneden2011} found similar $r$-process dispersion in M92, another ancient, metal-poor cluster, with~elements like Y, Zr, La, and~Eu showing clear star-to-star variations. This raised the possibility that such signatures might be a common feature of the most ancient GCs. Such signatures were also found in globular cluster escapees \citep{bandyopadhyay2}. Yet, not all studies agreed.  Ref. \citep{cohen2011}, using high-resolution Keck/HIRES data, failed to confirm the M92 dispersions, highlighting the delicate interplay between data quality and astrophysical interpretation. Ref. \citep{Roethompson2015} further cautioned that small systematic errors in atmospheric parameters could mimic abundance~spreads.

The past two years have ushered in a resurgence of clarity. Ref. \citep{kirby2023}, for~the first time, provided robust evidence for $r$-process dispersion in M92, but~with a remarkable twist: the variation was largely confined to first-generation stars (low-Na and high-Mg), marking the first observed link between light-element and neutron-capture abundances in a GC. Their findings imply a localized, inhomogeneous injection of $r$-process material during the cluster’s earliest formation phase, preceding the formation of second-generation stars by at least 0.8 Myr. Meanwhile, Ref. \citep{cabreragarcia2024} presented an extensive spectroscopic study of M15, finding that a majority of its stars are either $r$-I or $r$-II stars, and~confirmed significant dispersion in the neutron-capture elements from Sr to Dy. Intriguingly, they also uncovered tentative correlations between Na and $r$-process abundances\hl{---}an unexpected result that hints at a more complex chemical evolution than previously thought. Adding to this, Ref.~\citep{Ban2025} analyzed NGC 2298 and~found significant Sr and Eu dispersions among first-generation stars, along with correlations between [Sr/Eu], [Ba/Eu], and~[Mg/Fe], suggesting a shared origin for light $r$-process elements and Mg. The~cluster follows a universal $r$-process pattern, but~with greater scatter for main $r$-process elements than for the limited-$r$ group, highlighting multiple early enrichment~pathways.

Collectively, these discoveries transformed the narrative of GCs from simple relics to dynamic archives of early cosmic nucleosynthesis. The~evidence for real and intrinsic $r$-process dispersion in clusters like M15, M92, and~NGC 2298 suggests that some GCs were enriched by rare, high-yield events such as NSMs or exotic supernovae, while others were not, highlighting the stochastic nature of $r$-process enrichment in the early Universe. These ancient stellar systems thus offer a powerful probe into the timing, frequency, and~diversity of heavy-element production in the Galaxy's infancy. Moving forward, detailed chemical tagging, precise abundance measurements, and~next-generation simulations are essential to map the fingerprints of the earliest $r$-process events and their role in shaping the chemical landscape of the Milky~Way.

\subsection{Galactic~Stars}

Individual Galactic stars with over-abundances of $r$-process elements have been recognized for several decades (starting with~\citep{Griffin1982}) and~have been reported in the literature on a case-by-case basis ever since. However, until~recently, there has not been a sufficiently large sample of RPE stars to make significant headway on understanding the nature of their progenitor(s). Fortunately, dedicated surveys to identify larger inventories of stars with elemental-abundance patterns associated with the $r$-process have been underway over the course of the past decade. These include the RPA, mentioned above, as~well as detailed observations from the Chemical Evolution of R-process Elements in Stars (CERES) effort~(\citep{puls2025} and~references therein). The~upcoming RPA release of the results of high-resolution spectroscopic follow-up of some 2000 metal-poor stars, expected to include numerous new limited-$r$, $r$-I, and~$r$-II stars, will expand their numbers by roughly a factor of at least two and~hopefully enable better~understanding.

\begin{figure}[H]
\includegraphics[scale=0.9]{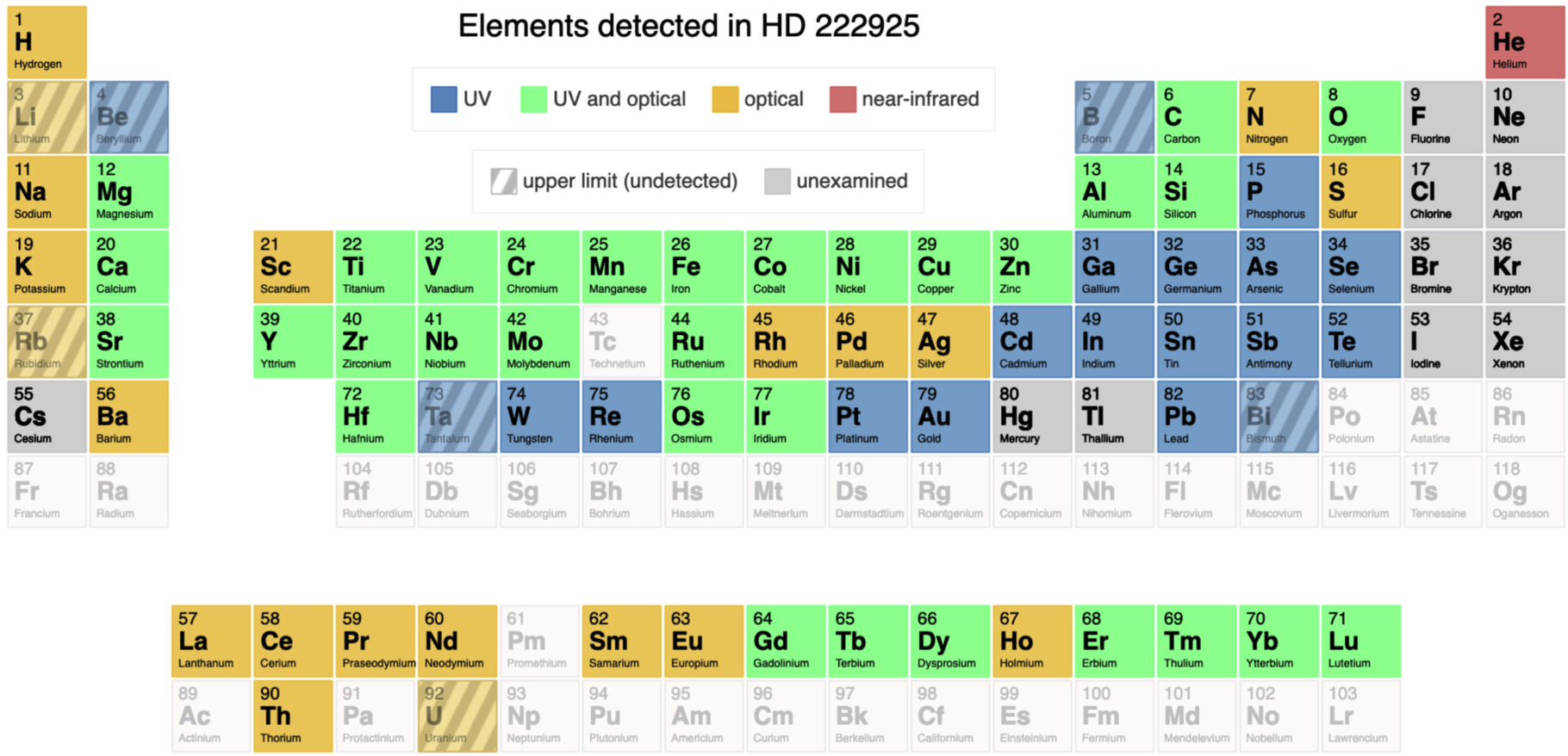}
\caption{\hl{This} 
figure, taken from~\citep{roederer2022}, shows the elements that have been detected and measured in HD 222925 using UV and optical spectroscopy. This also illustrates the importance of UV spectroscopy to maximize the number of elements studied in a star and~hence derive constraints on its production sites and progenitors. This figure is reproduced with permission © AAS and was originally published in \textit{The Astrophysical~Journal}.}
\label{HDstar}
\end{figure}

Observations of individual Galactic stars have clear advantages for the detailed study of the $r$-process, as~they are significantly brighter than stars in dwarf galaxies and globular clusters and~allow for the examination of main-sequence dwarfs rather than exclusively giants. The~enables high-S/N spectra to be obtained, from~which even weak lines of neutron-capture elements can be examined. In~addition, a~subset of such stars are sufficiently bright that they can be observed at high resolution in the near-UV with HST/STIS, opening pathways to include elements that do not possess lines in the optical range. The~powerful combination of ground-based and space-based high-resolution observations is clearly shown by the case of the halo star HD~222925, a~metal-poor star with the most complete set of abundances for any object outside the Solar System (\citep{roederer2022}; see Figure~\ref{HDstar}).

With the advent of the Gaia mission~\citep{gaia}, precision astrometry (geometric parallaxes and proper motions), and~for brighter stars, moderate-resolution spectroscopy in the region of the Ca triplet, from~which radial velocities can be obtained, have enabled the derivation of dynamical parameters for individual RPE stars. Based on this information, a~growing number of studies (e.g.,~\citep{roederer18_spvel,gudin2021,hattori2023,shank2023}) have explored the dynamical associations of such stars, tracing them back to their likely sites of origin (expected to be primarily disrupted dwarf galaxies). These so-called ``chemo-dynamically tagged groups'' (CDTGs) allow inferences to be made of the range of masses of their parent dwarf galaxies and~provide information on their chemical-evolution histories. In~this regard, it is notable that it has been recently claimed that a large fraction (at least $\sim$20\%) of all presently recognized RPE stars in the Galactic halo might be stars that were stripped from the Ret II dwarf galaxy~\citep{berczik2024}.

It is also remarkable that a few RPE stars have been identified recently at high metallicity
[Fe/H] $> -1.0$, which appear to be members of the disk system of the MW~\citep{Xie2024}. These include a thin-disk star (LAMOST J020623.21+494127.9), with~metallicity \mbox{[Fe/H] = $-0.54$} and [Eu/Fe] = +1.32, that possesses the highest reported Eu to Fe ratio ([Eu/H] = +0.78) known. Although~the study of RPE stars that are members of the disk system is still in its infancy, they have the potential to constrain its formation history, and~the conditions under which RPE stars can form at high~metallicity.

{While nearly all current observational constraints on the $r$-process are based on elemental abundances, isotopic abundances, if~measurable, could provide a transformative window into nucleosynthesis processes. Such data could, for~example, distinguish between different $r$-process pathways (main vs. weak), test predictions of fission-fragment distributions, or~reveal signatures of multiple enrichment events. However, measuring isotopic abundances in stars is extremely challenging due to limited spectral resolution, blending, and~the small isotopic shifts in most transitions. For~a few bright Galactic stars, isotopic ratios have been measured for elements like Ba and Eu, primarily using hyperfine splitting and isotope shifts in high-resolution spectra (e.g.,~\citep{Sneden2002, Aoki2003,Gallagher2012}). Future advances in instrumentation and modeling, 
especially in the infrared and UV spectra, may improve the feasibility of such measurements.}

\section{Looking~Ahead}

There are a number of ``mega'' spectroscopic surveys now underway or planned to start in the near future.
These include the Milky Way Survey within DESI~\citep{cooper2023}, the~4-metre Multi-Object Spectroscopic Telescope survey (4MOST;~\citep{dejong2019}), and~sub-surveys within the wide-field, massively multiplexed spectroscopic survey facility for the William Herschel Telescope (WEAVE;~\citep{jin2024}), all of which target individual Galactic stars, at~medium to high spectral resolution. The~Subaru Prime Focus Spectrograph survey (PFS;~\citep{takada2014}) will target the regions of Northern Hemisphere dwarf galaxies and~of course will include foreground/background stars in their neighborhood. A~high-resolution study of individual stars in Southern Hemisphere dwarf galaxies is presently being carried out with Gemini-S/GHOST. The~Multi-Object Optical and Near-infrared
Spectrograph (MOONS;~\citep{Gonzales2020}) survey on the VLT will provide moderate ($R \sim 5000$ and $R \sim 20,000$) optical and near-IR spectroscopy for stars in the Galactic Bulge, as~well as in the disk~system.

All of the above spectroscopic surveys will contribute to the identification of stars with over-abundances of neutron-capture elements, so we can expect significant numbers of limited-$r$, $r$-I, and~$r$-II (as well as numerous CEMP-$r$, CEMP-$s$, and~CEMP-$r/s$) stars to be found. The~final Gaia data release, planned for roughly two years hence, will contribute hundreds of millions of additional Galactic stars with available astrometry, radial velocities, and~very low-resolution spectra (the so-called XP spectra). For~optimal interpretation, many of the stars identified by these surveys will need to be observed at higher resolution and/or S/N. The~combination of precise astrometry and radial velocities from Gaia will enable an enormous expansion of the numbers of stars for which CDTGs can be searched, helping to constraint the nature of their~birthplaces.

Given their relative rarity, the~need to identify substantially larger samples of, in~particular,
VMP, EMP, and~UMP stars in order to explore the nature of the $r$-process at the lowest metallicities should be given high priority. Mega photometric surveys involving narrow- and medium-band filters, mentioned below, are an ideal method to accomplish this. Over~the past few years, intensive efforts have been made to refine the photometric zero-points for such surveys and~for the derivation of stellar parameters and elemental abundances~\citep{Huang2021,Huang2022,Xu2022,Yang2022,Huang2023,
Xiao2023,Huang2024,Lu2024,Xiao2024,Gu2025,Huang2025,Zhang2025}.
The combination of such filters with ultra-wide Gaia $BP$ and $RP$ photometry has been shown to provide accurate estimates of atmospheric parameters, including metallicities, and~in some cases, individual elemental abundances (e.g., [C/Fe], [Mg/Fe], and [$\alpha$/Fe]) for stars down to about [Fe/H] = $-4.0$.

The Stellar Abundance and Galactic Evolution Survey (SAGES;~\citep{Zheng2018,Fan2023}) and the SkyMapper Southern Survey (SMSS;~\citep{Keller2007,Onken2024}) efforts have been completed. The~Southern Photometric Local Universe Survey (S-PLUS;~\citep{Mendes2019}) and the Javalambre Photometric Local Universe Survey (J-PLUS;~\citep{Cenarro2019}) are ongoing and~are expected to be completed in the next few years. The~PRISTINE survey~\citep{Starkenburg2017} is also ongoing. These surveys have already produced long lists of candidate EMP and UMP stars suitable for moderate-to-high-resolution follow-up, from~which neutron-capture enhanced stars can be identified. The~estimated numbers of VMP/EMP/UMP stars expected to be found based on the combination of the photometriċ surveys to date are prodigious: tens of millions of stars with [Fe/H] $\leq -2.0$, hundreds of thousands with [Fe/H] $\leq -3.0$, and~several thousand stars with [Fe/H] $\leq -4.0$.

The Ultra Short Survey (USS;~\citep{Perottoni2024}), a~sub-survey within S-PLUS, is of particular note, as~it is based on substantially shorter exposure times than the normal S-PLUS program. This enables much brighter stars to be observed without saturation, including stars as bright as $G \sim 7.0$.
The Javalambre Physics of the Accelerating Universe Astrophysical Survey (J-PAS;~\citep{Bonoli2021}), already underway, is based on over 50 narrow-to-medium-band filters over most of the optical region. Based on these data, it is expected that numerous individual elemental abundances can be estimated, possibly including several neutron-capture~elements.

An auxiliary survey, the Mapping the Ancient Galaxy In CaHK (MAGIC) survey, using a narrow-band CaII HK filter in combination with observations from the already completed Dark Energy Survey, which included only broad-band filters, is underway with the Dark Energy Camera (DECam) at the 4 m Blanco Telescope (see~\citep{Barbosa2025}). Similar efforts in order to supplement the soon-to-begin Legacy Survey of Space and Time (LSST; see~\citep{LSST2009,Ivezic2019}) at the Vera C. Rubin Observatory would also be of great~interest.

{A major challenge in modeling $r$-process nucleosynthesis arises from the scarcity of experimental nuclear data for the very neutron-rich nuclei involved. Many of these isotopes lie far from stability and have extremely short half-lives, making them inaccessible to direct measurements with current accelerator facilities. As~a result, key inputs such as nuclear masses, $\beta$-decay rates, neutron-capture rates, and~fission yields often rely on theoretical models, which introduce significant uncertainties in the predicted abundance patterns. Efforts at next-generation radioactive ion beam facilities (e.g., FRIB, FAIR, and RIKEN) are beginning to extend the reach of experiments toward the relevant regions of the nuclear chart, but~comprehensive coverage remains a long-term goal. These experimental limitations underline the need for close collaboration between nuclear theory, laboratory measurements, and~astrophysical observations.}

{Finally, it must be acknowledged that in~$r$-process nucleosynthesis the~resulting abundance patterns are strongly influenced by the underlying nuclear physics inputs. Although~$r$-process elements are mainly produced through rapid neutron capture and beta decay, the~uncertainties due to specific reaction rates and cross sections are comparatively low under the extreme conditions of high temperature and neutron density. Instead, nuclear masses (which determine (n,$\gamma$) Q-values) and $\beta$-decay rates along the neutron-rich isotopic chains play a dominant role in shaping the abundance pattern and setting the timescales. Additionally, fission processes, including fission recycling, can significantly alter the production of the heaviest nuclei, particularly in very neutron-rich environments such as neutron star mergers. Recent progress in experimental and theoretical nuclear physics such as measurements of $\beta$-decay half-lives at radioactive ion beam facilities and improved global mass models has begun to refine our understanding of these processes. These nuclear properties are often 
uncertain and far from stability, and~current experimental and theoretical efforts are helping to reduce these uncertainties as discussed in~\citep{Mum2016,thielemann2017,Horowitz2019}.
}

\vspace{6pt}

\authorcontributions{Conceptualization, A.B. and T.C.B.; methodology, A.B. and T.C.B.; validation, A.B. and T.C.B.; formal analysis, A.B. and T.C.B.; investigation, A.B. and T.C.B.; resources A.B. and T.C.B.; writing---original draft preparation, A.B. and T.C.B.; writing---review and editing, A.B. and T.C.B.; visualization, A.B.; supervision, T.C.B.; project administration, A.B. and T.C.B. All authors have read and agreed to the published version of the manuscript.}

\funding{T.C.B. acknowledges funding support from grant PHY 14-30152: Physics Frontier Center/JINA Center for the Evolution of the Elements (JINA-CEE), and from OISE-1927130: The International Research Network for Nuclear Astrophysics (IReNA), awarded by the US National Science~Foundation.}

\conflictsofinterest{The authors declare no conflicts of interest.}

\begin{adjustwidth}{-\extralength}{0cm}
\printendnotes[custom]
\reftitle{References}



%


%
%
%
\PublishersNote{}
\end{adjustwidth}
\end{document}